\documentclass[twocolumn,twoside]{revtex4}
\usepackage{graphicx}
\usepackage{bm}
\usepackage{amssymb}
\usepackage{fancyhdr}
\usepackage{epstopdf}
\usepackage{textpos}
\newcommand{\sign}{\mathop{\mathrm{sign}}\nolimits}

\begin{document}

\title{\centering\boldmath Review of \( B_s \) Mesons and \( b \) Baryons }


\author{
\centering
\begin{center}
R.F.~Harr
\end{center}}
\affiliation{\centering Wayne State University, Detroit Michigan, 48202, USA}
\begin{abstract}
The measurements of \(B_s^0\) mesons and \(b\) baryons advanced greatly in the past year.
The ground state \( b \) baryon \( \Xi_b^0 \) was observed for the first time; numerous decay modes of the \( B_s^0 \) and \( b \) baryons were newly seen; measurements of CP violation in \( B_s^0 \) decays were improved, generally improving the agreement with the standard model; and the search for rare decays sensitive to new physics were pushed to new levels.
It was a very productive year for measurements of \(b \) baryons and \( B_s^0\) mesons.
\end{abstract}

\maketitle
\thispagestyle{fancy}


The majority of the measurements of \( B_s \) mesons and \( b \) baryons have come from hadron colliders.
During the past 20 years the Tevatron experiments, CDF and D0, have provided the bulk of the measurements, including the measurement of the \( B_s \) mixing frequency, the initial measurements of the \( B_s \) mixing phase and lifetime difference, and observation of all but one of the ground state \( b \) baryons.
The LHC experiments are now taking the lead in producing new results as demonstrated by those available in Summer 2011.

The B factories provide important information about the \( B_s \) mesons:  reference modes for measurements at hadron colliders, and measurements difficult to make at hadron colliders.

\section{ Results for \boldmath\( b \)  Baryons }

The initial observations of \( b \) baryons were made by UA1, the split-field magnet, and LEP experiments.
These were limited to observations of the \( \Lambda_b^0 \) baryon, a few decay modes, and evidence for other ground state \( b\) baryons from semileptonic decays.
The Tevatron experiments provided a large number of measurements including direct observations of most of the ground state \( b \) baryons, precision lifetime measurements, and observations of additional decay modes.
Now results from the LHC experiments are also arriving, and with the expected size of the data set, the possibility of many new measurements coming in the near future.

Despite two decades of measurements, the actual amount known about \( b \) baryons is rather sparse.
For most of the \( b \) baryons, one or two decay modes have been seen, and only about a dozen are known for the \( \Lambda_b^0 \).

\subsection{\boldmath Observation of \( \Xi_b^0 \) }

The observation of the \( \Xi_b^0 \) by CDF involves a rather complicated decay topology with 6 final state tracks and 4 decay vertices \cite{Cascade}.
The reconstructed decay sequence is \( \Xi_b^0 \to \Xi_c^+ \pi^- \) followed by \( \Xi_c^+ \to \Xi^- \pi^+ \pi^+ \), \( \Xi^- \to \Lambda \pi^- \), and \( \Lambda \to p \pi^- \).
The similar decay sequence is also reconstructed \( \Xi_b^- \to \Xi_c^0 \pi^- \).

\begin{figure}
\includegraphics[width=75mm]{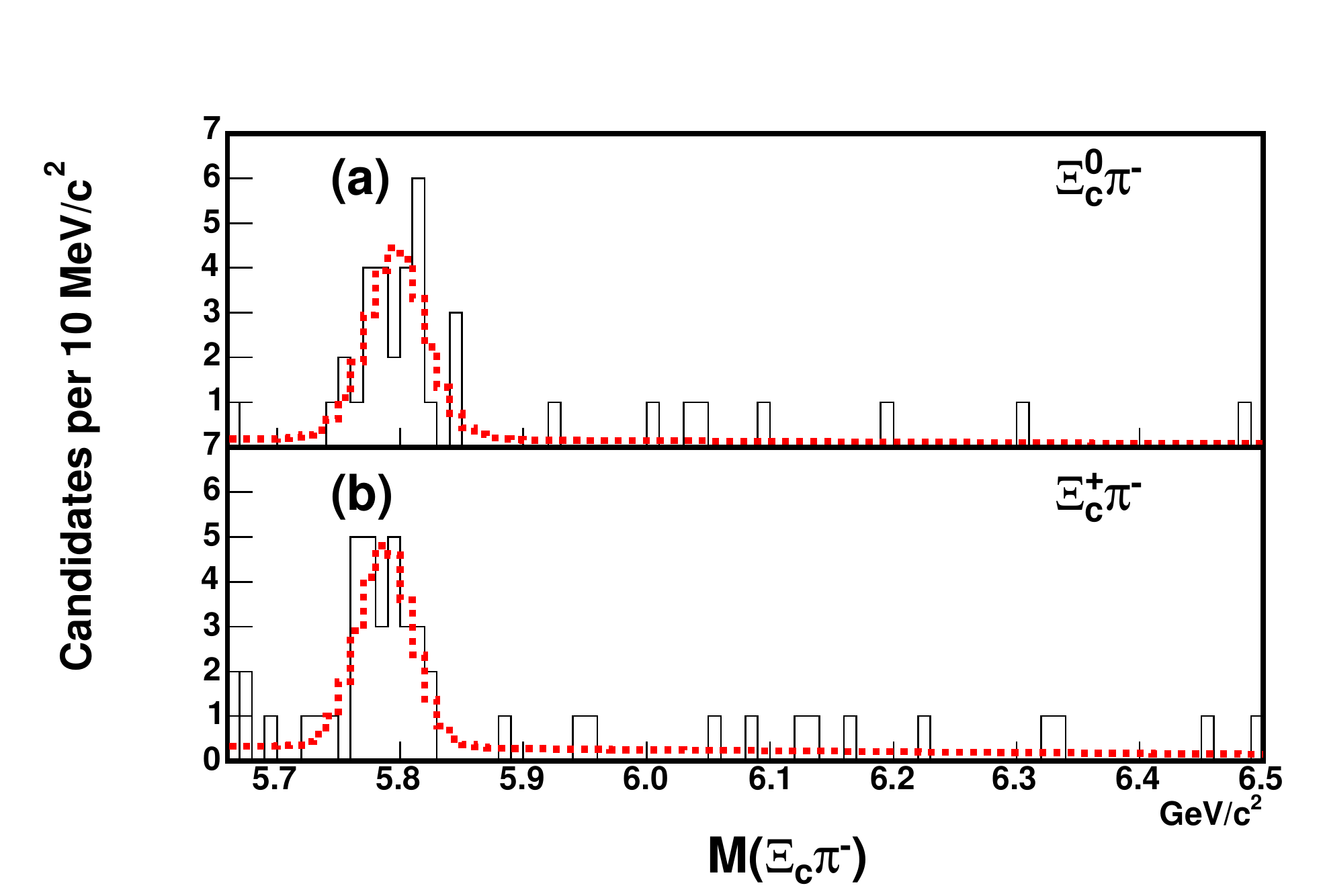}
\caption{The invariant mass of the (a) \( \Xi_b^- \) reconstruction and (b) \( Xi_b^0 \) reconstruction.
The region below about \( 5.7\,\mathrm{GeV/}c^2\) is populated with partially reconstructed decays and is not used for evaluating the existence of a signal. }
\label{fig:Cascade_b}
\end{figure}

\subsection{\boldmath \( \Lambda_b^0 \) Production Cross Section and \( \mathcal{B}\left( \Lambda_b^0 \to J/\psi \Lambda \right) \) }

Decays of \( b \) baryons to a \( J/\psi \) are relied on for many measurements.
The ATLAS and CMS experiments report observations of the decay \( \Lambda_b^0 \to J/\psi \Lambda \) \cite{ATLAS_LbtoJpsiL, CMS_LbtoJpsiL}.
ATLAS observes \( 579 \pm 31 \) events in \( 1.2\,\mathrm{fb}^{-1} \), while CMS find about 100 events in \( 194 \,\mathrm{pb}^{-1} \).
Both collaborations are exploring possibilities to make mass, lifetime, branching fraction, and other measurements.

Quark fragmentation is a non-perturbative process that must be modeled in simulations.
One of the pieces that must be modeled is the probability that the fragmentation of a quark will yield a baryon.
The D0 collaboration has used the decay \( \Lambda_b^0 \to J/\psi \Lambda \) to determine the branching fraction times the probability that a \( b \) quark will fragment to a \( \Lambda_b^0 \) baryon \cite{D0_LbtoJpsiL}.
They find
\[ f(b \to \Lambda_b) \mathcal{B}( \Lambda_b^0 \to J/\psi \Lambda ) = (6.01 \pm 0.88) \times 10^{-5}. \]

\subsection{ \boldmath Newly Seen \( \Lambda_b^0 \) Decay Modes  }

\begin{figure*}
\includegraphics[width=150mm]{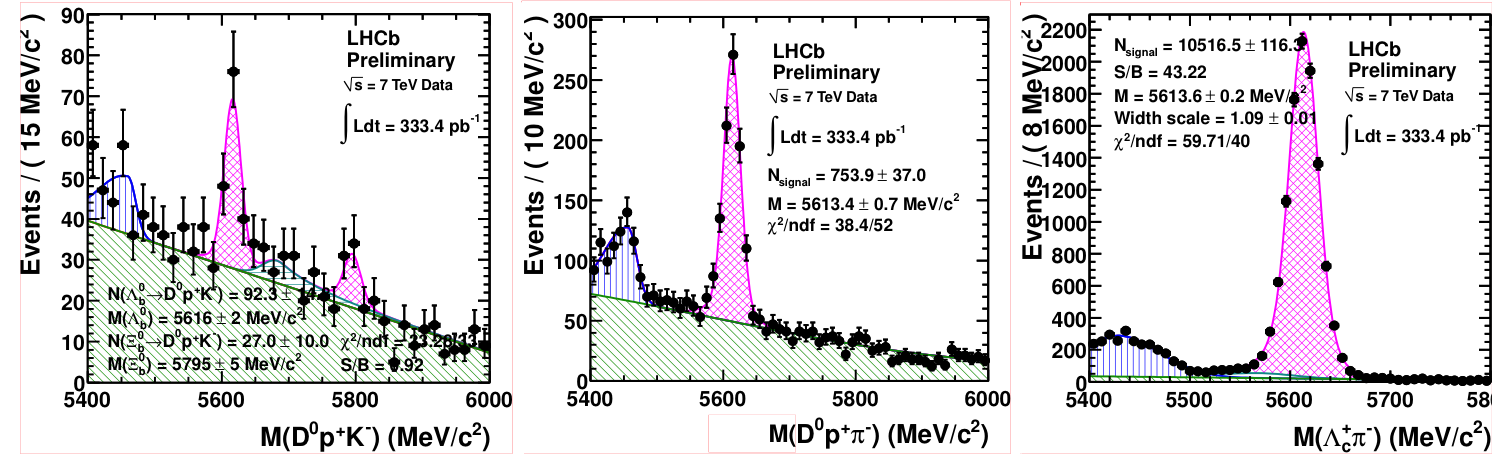}
\caption{The LHCb experiment reconstructed the \( \Lambda_b^0 \) with \( 333.4 \,\mathrm{pb}^{-1} \) of integrated luminosity in three decay modes: \( D^0 p K^- \) (left), \( D^0 p \pi^- \) (center), and \( \Lambda_c^- \pi^+ \) (right).
The yields and mass values are displayed on the plots.
Evidence for the decay \( \Xi_b^0 \to D^0 p K^- \) is seen in the left plot as the peak near \( 5800 \,\mathrm{MeV}/c^2 \).
}
\label{fig:LHCb_LbFits}
\end{figure*}

A number of \( \Lambda_b^0 \) decay modes were newly seen this year.
The LHCb collaboration \cite{LHCb_LbFits} reported seeing \( 92 \pm 14 \)  decays \( \Lambda_b^0 \to D^0 p K^- \), a previously unobserved decay mode (see Fig.~\ref{fig:LHCb_LbFits}).
The same mass fit contains evidence for the recently observed state \( \Xi_b^0 \) decaying to the same final state.
They also reconstruct the \( \Lambda_b^0 \) in two additional decay modes, \( D^0 p \pi^- \) and \( \Lambda_c^- \pi^+ \).

The CDF collaboration reported the first observation of the flavor--changing neutral--current decay \( \Lambda_b^0 \to \Lambda \mu^+\mu^- \) \cite{CDF_LbtoLmumu}.
This decay is sensitive to new physics in the differential branching fraction and asymmetries in angular distributions.
CDF reported a signal of \( 24 \pm 5 \) decays with a significance of about 6 Gaussian sigmas.
The branching fraction is
\[ \mathcal{B}(\Lambda_b^0 \to \Lambda \mu^+\mu^- ) = [ 1.73 \pm 0.42 (\mathrm{stat}) \pm 0.55 (\mathrm{syst})] \times 10^{-6} \]
the smallest \( \Lambda_b^0 \) branching fraction yet measured.

\section{\boldmath Results for \( B_s^0 \) Mesons }

The \( B_s^0 \) meson is much better studied than the \(b \) baryons, but still far short of the well--studied \( B^0 \) and \( B^+ \) mesons.
Studies are carried out at \( e^+e^- \) colliders in \( \Upsilon(5S) \) decays, but the greatest numbers are produced at hadronic colliders.
This fact is exemplified by the observation of \( B_s^0 \) mixing in 2005 by the CDF collaboration.

Recently, there has been great interest in measurements sensitive to new physics, either through the \( B_s \) mixing phase, or the rare decay \( B_s \to \mu^+\mu^- \).
New measurements were released by a number of experiments.
These results are discussed in the following.

\subsection{\boldmath Update on Measurement of the Mixing Phase \( \phi_s \) and Lifetime Difference \( \Delta \Gamma_s \) }

\subsubsection{\boldmath Dimuon Charge Asymmetry \unboldmath}

The D0 collaboration updated the dimuon charge asymmetry measurement to \( 9.0\,\mathrm{fb}^{-1} \) of integrated luminosity.
The asymmetry is the difference in the number of events with a pair of positive muons minus the number with a pair of negative muons divided by the sum.
The like sign pairs come predominately from events with a \( b \bar{b} \) pair of quarks, hadronizing to yield a \( b \) hadron and a \( B^0 \) or \( B_s^0 \) meson that oscillates to its anti-particle state, then both the \( b \) hadron and oscillated meson decay semi-muonically.
The standard model predicts a tiny asymmetry in the number of positive versus negative dimuon events.
A deviation from the standard model value could indicate new physics entering in the mixing of \( B \) mesons.

\begin{figure}
\includegraphics[width=70mm]{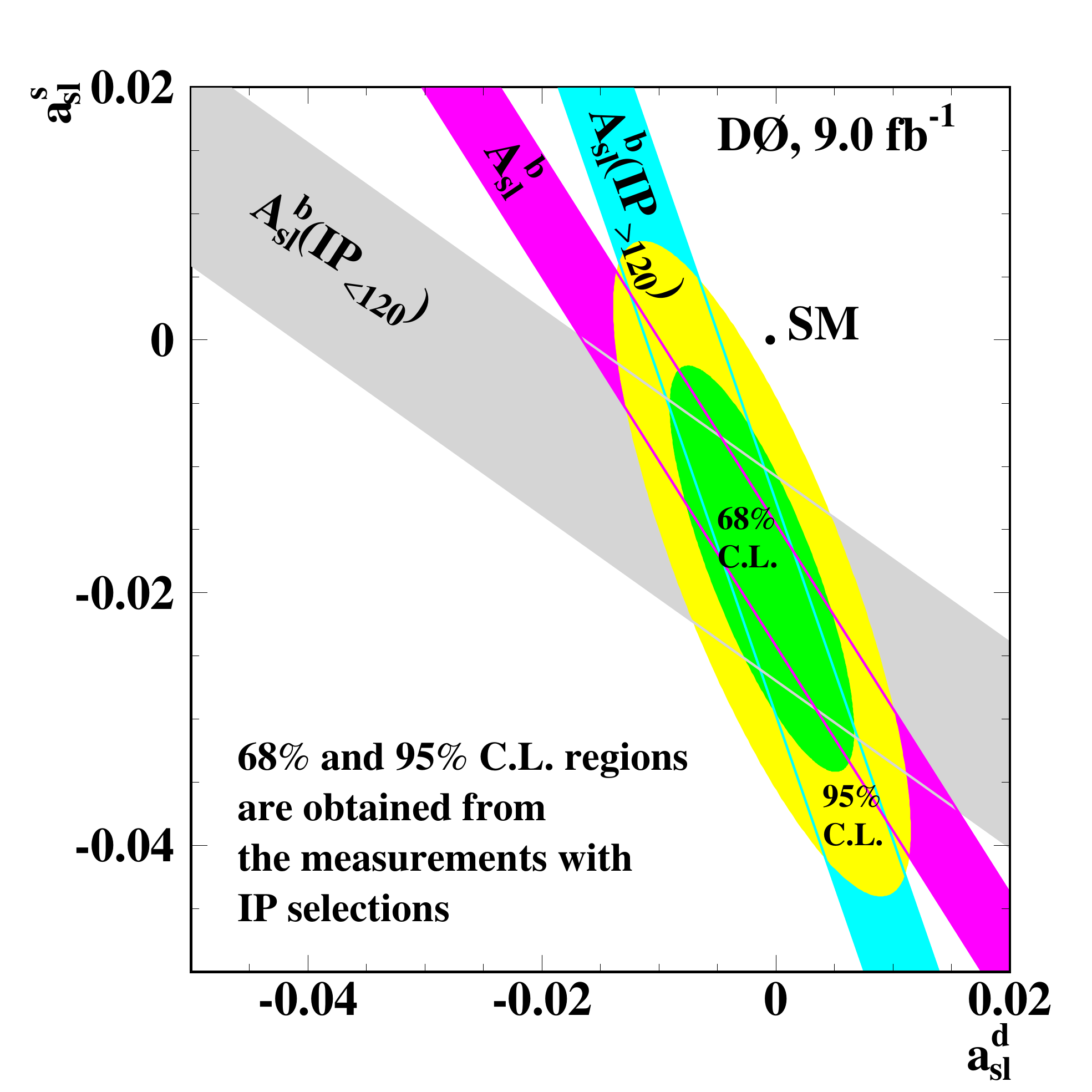}
\caption{The dimuon asymmetry measured by D0 can be interpreted in terms of asymmetries arising in the decay of a \( B^0 \), \( a_{sl}^d \), or a \(B_s^0 \), \( a_{sl}^s \), as indicated by the magenta band.
With an impact parameter requirement, the sensitivity to \( B^0 \) (large impact parameter) versus \( B_s^0 \) (small impact parameter) varies, as shown by the cyan and gray regions.
The overlap of these two regions is indicated by the 68\% and 95\% contours.
}
\label{fig:D0_Asl}
\end{figure}

The D0 experiment presented a result in 2010 that differed from the SM by \( 3.2 \) standard deviations.
The updated result is  \cite{D0_Asl}
\[ A_{sl}^b = [ -0.787 \pm 0.172 (\mathrm{stat}) \pm 0.093 (\mathrm{syst}) ]\%.\]
The central value of the updated D0 result is closer to the SM prediction, but has smaller uncertainty and lies \( 3.9 \) standard deviations from the SM.
The D0 collaboration further investigates the effect by analyzing the dependence of the result on an impact parameter requirement.
Muons with large impact parameter are likely to arise from \( b \) hadrons that live longer than those with small impact parameter.
The mixing of \( B_s^0 \) mesons is much faster than \( B^0 \) mesons, so like sign pairs where a muon has large impact parameter are more likely to arise from \( B^0 \) decays, and vice versa.
This dependence is used to constrain the dependence of the dimuon asymmetry on the asymmetries in \( B^0 \) and \(B_s^0 \) decays, \( a_{sl}^d \) and \( a_{sl}^s \), respectively, shown graphically in Fig.~\ref{fig:D0_Asl}.

\subsubsection{\boldmath \( CP \) Violation in \( B_s^0 \to J/\psi \phi \) \unboldmath}

The SM expectation for \( CP \) violation in \( B_s^0 \to J/\psi \phi \) decays is extremely small, making it difficult to extract the CKM phase.
However it is then relatively easy  to look for the effects of new physics producing an enhancement to the phase.
The first results from CDF and D0 sparked great interest because each deviated from the SM expectation by about \( 2\sigma \).
This  spurred work by many collaborations to confirm the result.

The CMS collaboration has reconstructed the decay and used it to measure the \( B_s^0 \) production distributions in transverse momentum \( p_T \) and rapidity \( y \) \cite{CMS_LbtoJpsiL}.
This is a prelude to a CP violation measurement.

\begin{figure}
\includegraphics[width=70mm]{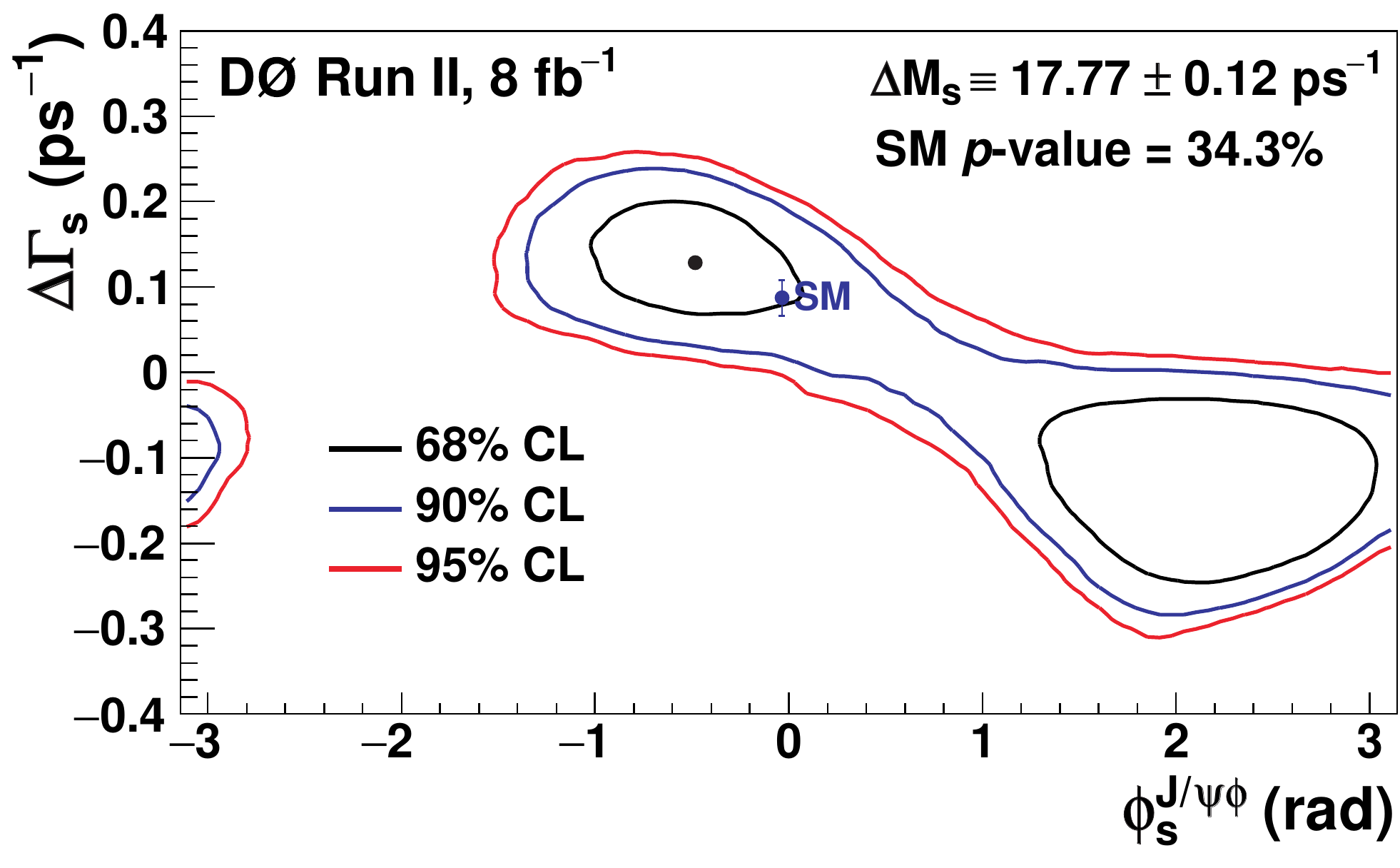}
\caption{Likelihood contours for \( \Delta \Gamma_s \) versus \( \phi_s \) determined in \( B_s^0 \to J/\psi \phi \) decays from the D0 experiment.
}
\label{fig:D0_Jpsiphi}
\end{figure}

The D0 experiment has updated their result for \( CP \) violation using \( 8\,\mathrm{fb}^{-1} \) of integrated luminosity \cite{D0_BstoJpsiphi}.
The new result lies about \( 1\sigma \) from the SM expectation, reducing the tension with the SM.

The LHCb experiment presented a first measurement of \( CP \) violation in \( B_s^0 \to J/\psi \phi \), summarized at this conference in the talk of P.~Koppenburg \cite{LHCb_bstoJpsiphi}.
Their result has smaller uncertainty than the D0 result and the earlier CDF result, but is consistent with both, and the SM expectation.
The initial possibility of a large contribution from new physics is fading, but we will need to await combinations of results from these experiments or even higher precision results to see if there is a measurable effect.

\subsection{\boldmath Newly Seen \( B_s^0 \) Decay Modes \unboldmath}

A significant number of \( B_s^0 \) decay modes were newly observed in the past year.
The observations have come from both \( e^+e^- \) and hadron colliiders, representing a major expansion of known decay modes.

\subsubsection{ \boldmath \( B_s^0 \to J/\psi f_0(980) \) \unboldmath}

The decay \( B_s^0 \to J/\psi f_0(980) \) is a pseudoscalar to vector -- scalar decay, and must proceed through \(S\)--wave channel.
This mode could contribute an \(S\)--wave component to the \( B_s \to J/\psi \phi \) analysis when the \( f_0 \) decays to \( K^+ K^- \).
The measurement of its branching fraction was desired to help constrain any effects in the \( J/\psi \phi \) analysis.
The final state is CP odd, so in the absence of CP violation in the decay, a measurement of its lifetime corresponds to the lifetime of the heavy \( B_s^0 \) state, \( \tau_H \).
The lifetime difference \( \Delta \Gamma_s \) can be determined from the average \( B_s \) lifetime, well measured in decays to non-CP eigenstates, and either of the light or heavy state lifetimes.
And the mixing phase \( \phi_s \) can be determined directly from decays to a CP eigenstate without the need for an angular analysis \cite{StoneZhang}.

Early in 2011, the LHCb collaboration released an observation of the \( B_s^0 \to J/\psi f_0(980) \) decay mode \cite{LHCb_BstoJpsif0}.
This was quickly followed by results from Belle \cite{BELLE_BstoJpsif0}, CDF \cite{CDF_BstoJpsif0}, and D0 \cite{D0_BstoJpsif0}.
The results of these measurements are all in general agreement and point to a ratio of the fraction of \( J/\psi f_0(980) \) decays to \( J/\psi \phi \) decays between about 1/5 and 1/3.
The LHCb collaboration also sees evidence for decays to other resonances that decay to \( \pi^+\pi^- \), in particular, the \( f'_2(1525) \).

The CDF collaboration goes on to measure the \(B_s^0 \) lifetime in this mode, the first \( B_s^0 \) lifetime measurement for a decay to a CP eigenstate.
The measurement is made by a simultaneous fit to the decay time distribution, mass distribution, and decay time error.
The result is
\[ \tau( B_s^0 \to J/\psi f_0(980) ) = 1.70 ^{+0.12}_{-0.11} (\mathrm{stat}) \pm 0.03 (\mathrm{syst})\,\mathrm{ps}. \]
When interpreted as \( \tau_H \), this result is in agreement with indirect determinations of \( \tau_H\) and theoretical predictions, albeit with a relatively large error.
The error is due, predominantly, to statistics.
Larger samples should be available in the near future from LHC experiments.
We should expect more precise measurements in the future, providing additional handles on the question of new physics effects in \( B_s^0 \) mixing.

\subsubsection{\boldmath \( B_s^0 \to J/\psi \eta \) \unboldmath}

\begin{figure}
\includegraphics[width=70mm]{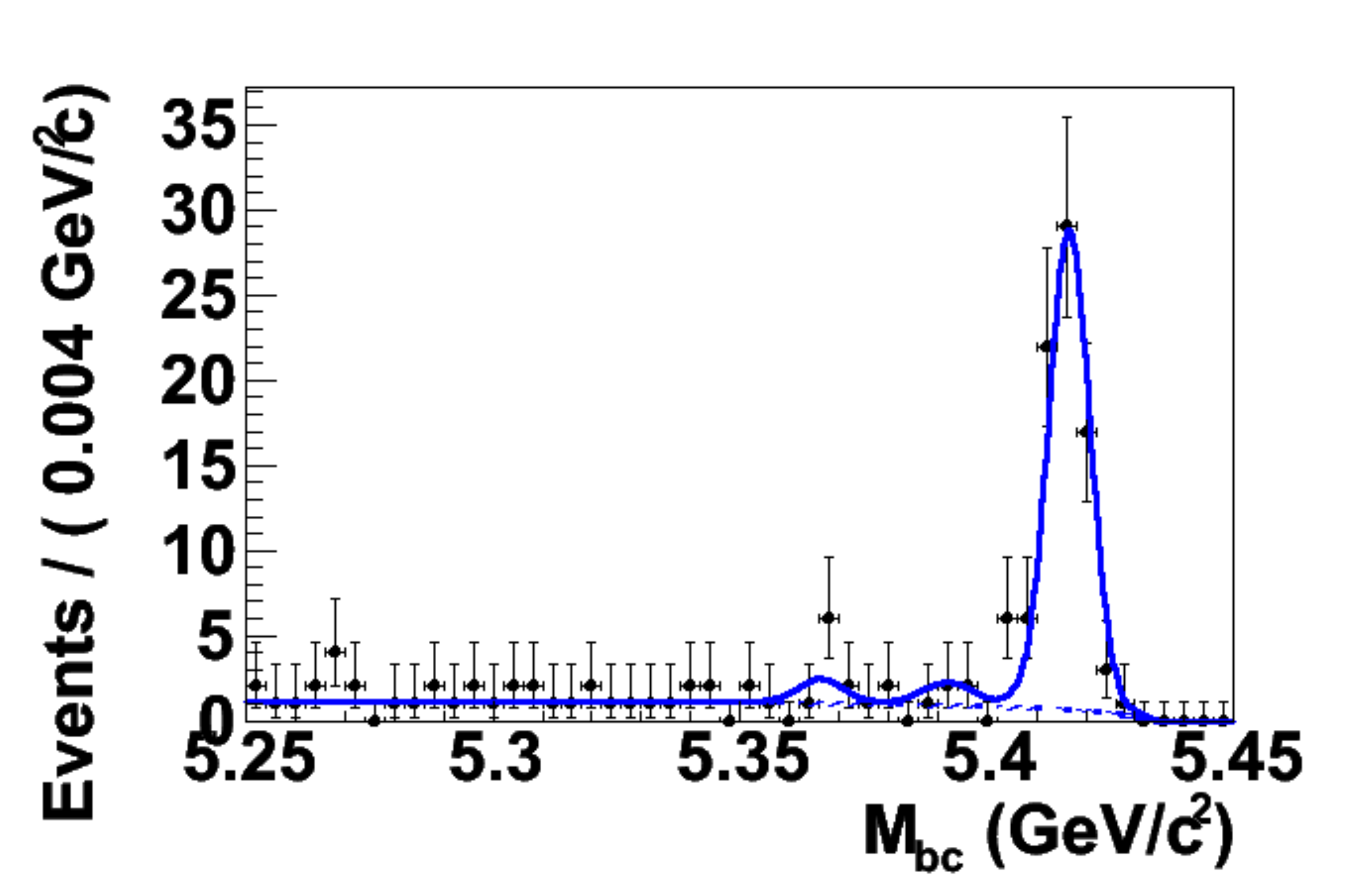} \newline
\includegraphics[width=70mm]{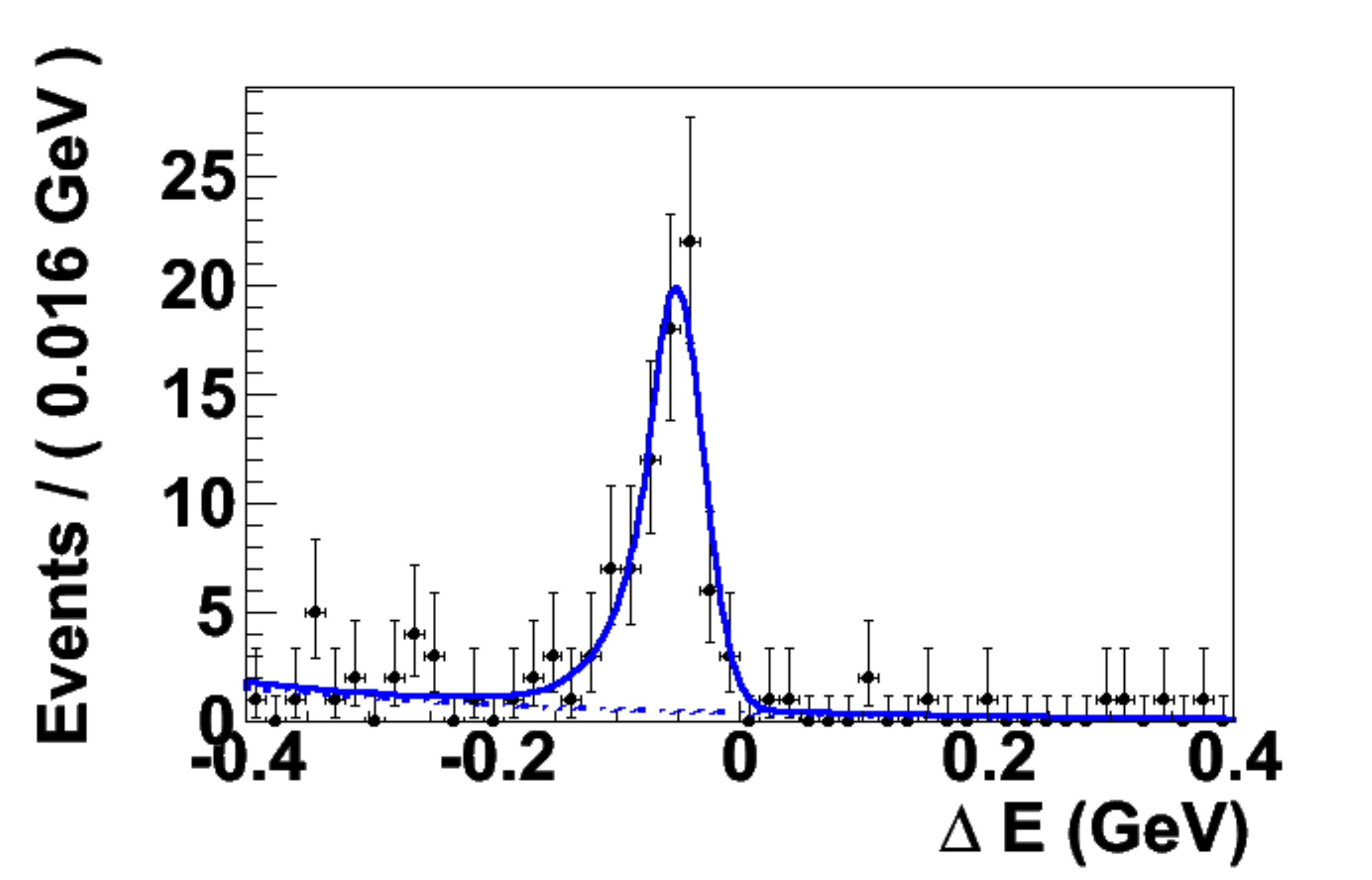}
\caption{The beam constrained mass \( M_{bc} \) distribution (top) and \( \Delta E \) distribution (bottom) for the \( B_s^0 \to J/\psi \eta \) candidates with \( \eta \) reconstructed in the \( \gamma\gamma \) mode.
Plots courtesy of the Belle experiment.
}
\label{fig:Belle_Jpsieta0}
\end{figure}

\begin{figure}
\includegraphics[width=70mm]{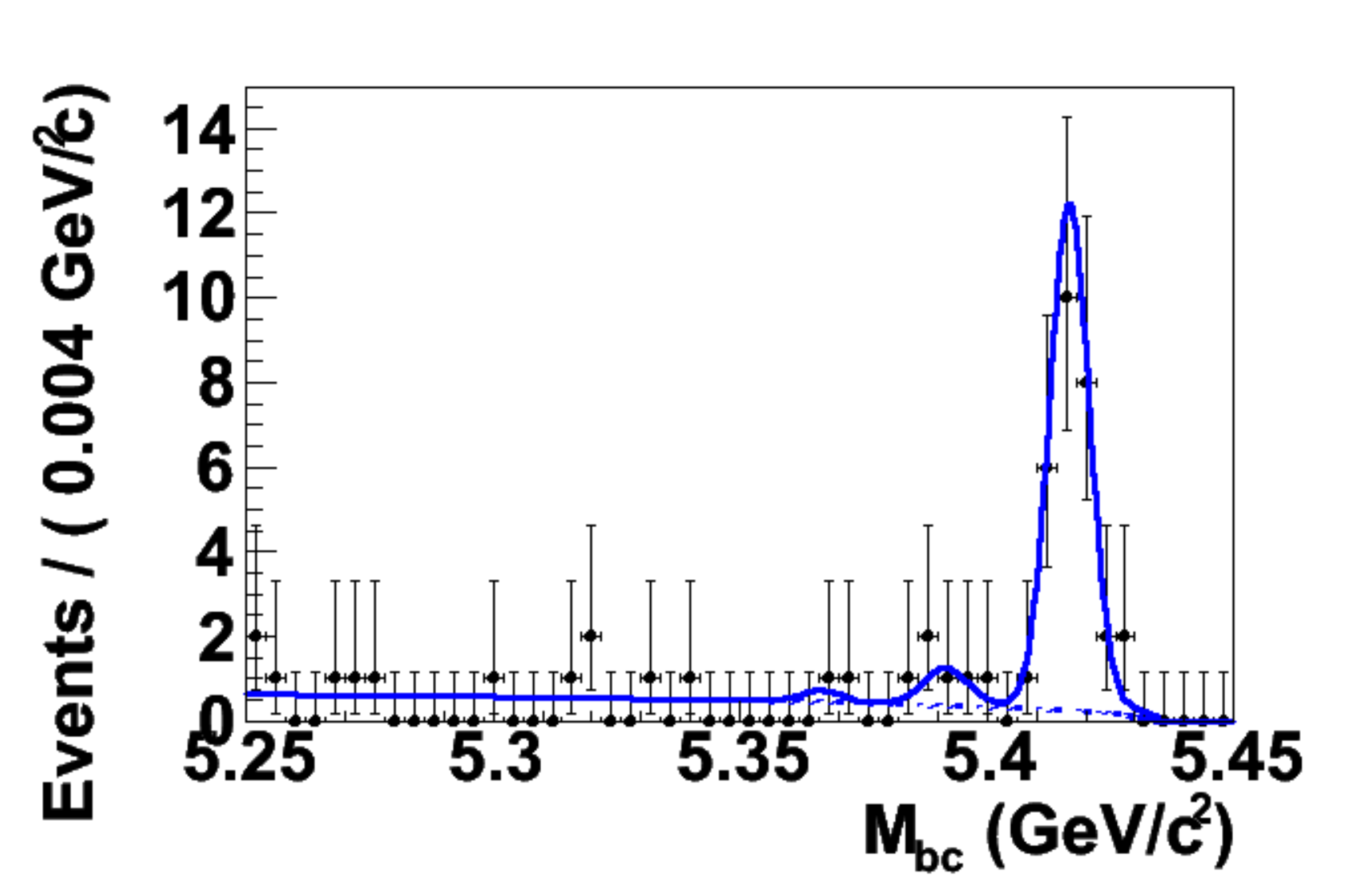} \newline
\includegraphics[width=70mm]{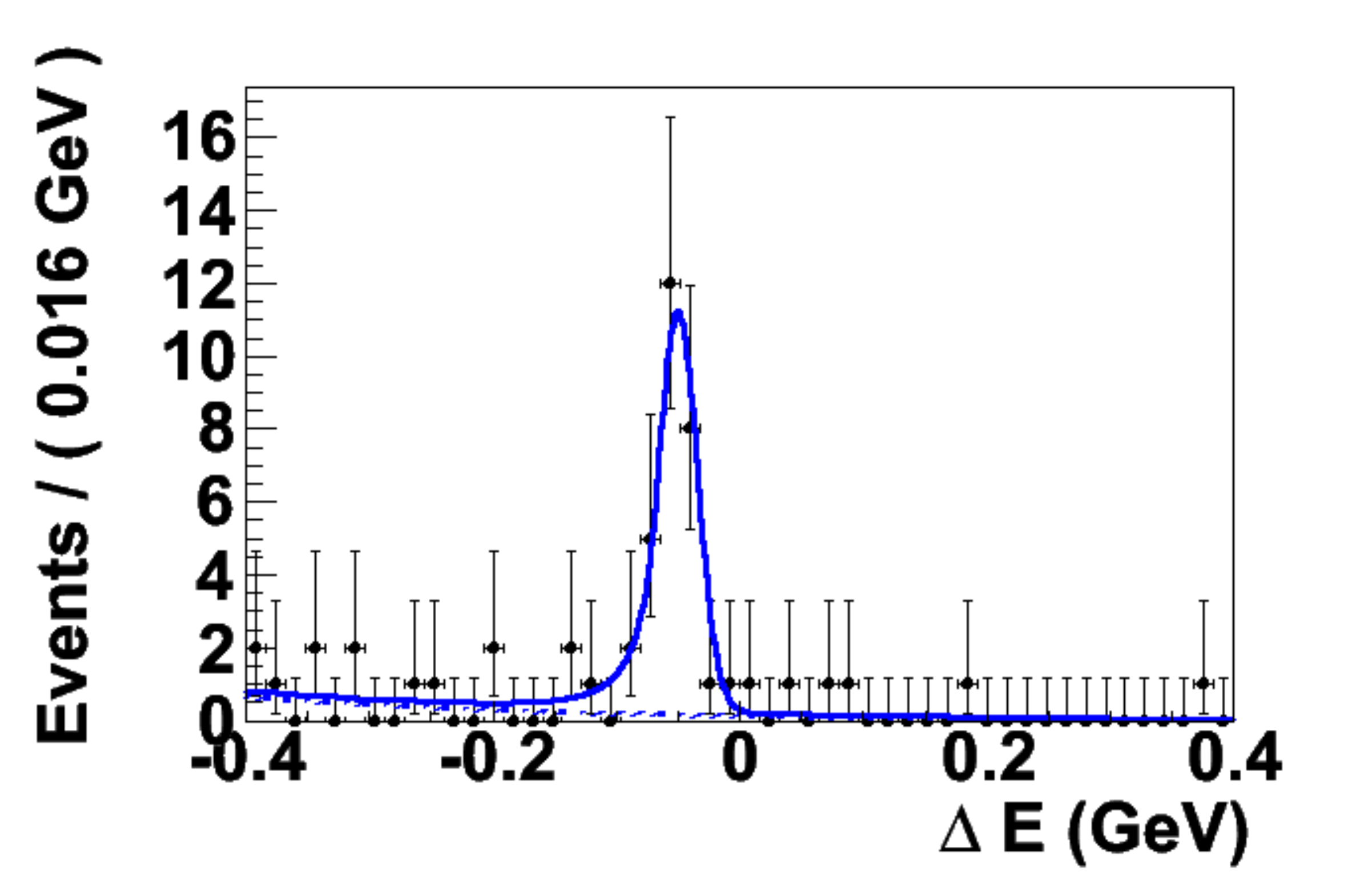}
\caption{The beam constrained mass \( M_{bc} \) distribution (top) and \( \Delta E \) distribution (bottom) for the \( B_s^0 \to J/\psi \eta \) candidates with \( \eta \) reconstructed in the \( \pi^+\pi^-\pi^0 \) mode.
Plots courtesy of the Belle experiment.
}
\label{fig:Belle_Jpsieta1}
\end{figure}

The Belle collaboration released a new result at this conference, the observation of the decay \( B_s^0 \to J/\psi \eta \), a CP  even final state.
The \( \eta \) is reconstructed in two decay modes, \( \gamma\gamma \) with \( 101 \pm 10 \) candidates and \( \pi^+\pi^-\pi^0 \) with \( 41 \pm 4 \) candidates.
The fits in \(M_{bc} \) and \( \Delta E \) are shown in Figs.~\ref{fig:Belle_Jpsieta0} and \ref{fig:Belle_Jpsieta1}.
The branching fraction is determined to be
\[ \mathcal{B}(B_s^0 \to J/\psi \eta) = \]
\[[5.11 \pm 0.50 (\mathrm{stat}) \pm 0.35 (\mathrm{syst}) \pm 0.68 (f_s)] \times 10^{-4}.\]
Since this is a decay to a CP eigenstate, in the absence of CP violation, a measurement of the lifetime can be used to determine \( \Delta \Gamma_s \).

\subsubsection{ \boldmath \( B_s \to \phi \mu^+ \mu^- \) \unboldmath}

\begin{figure}
\includegraphics[width=70mm]{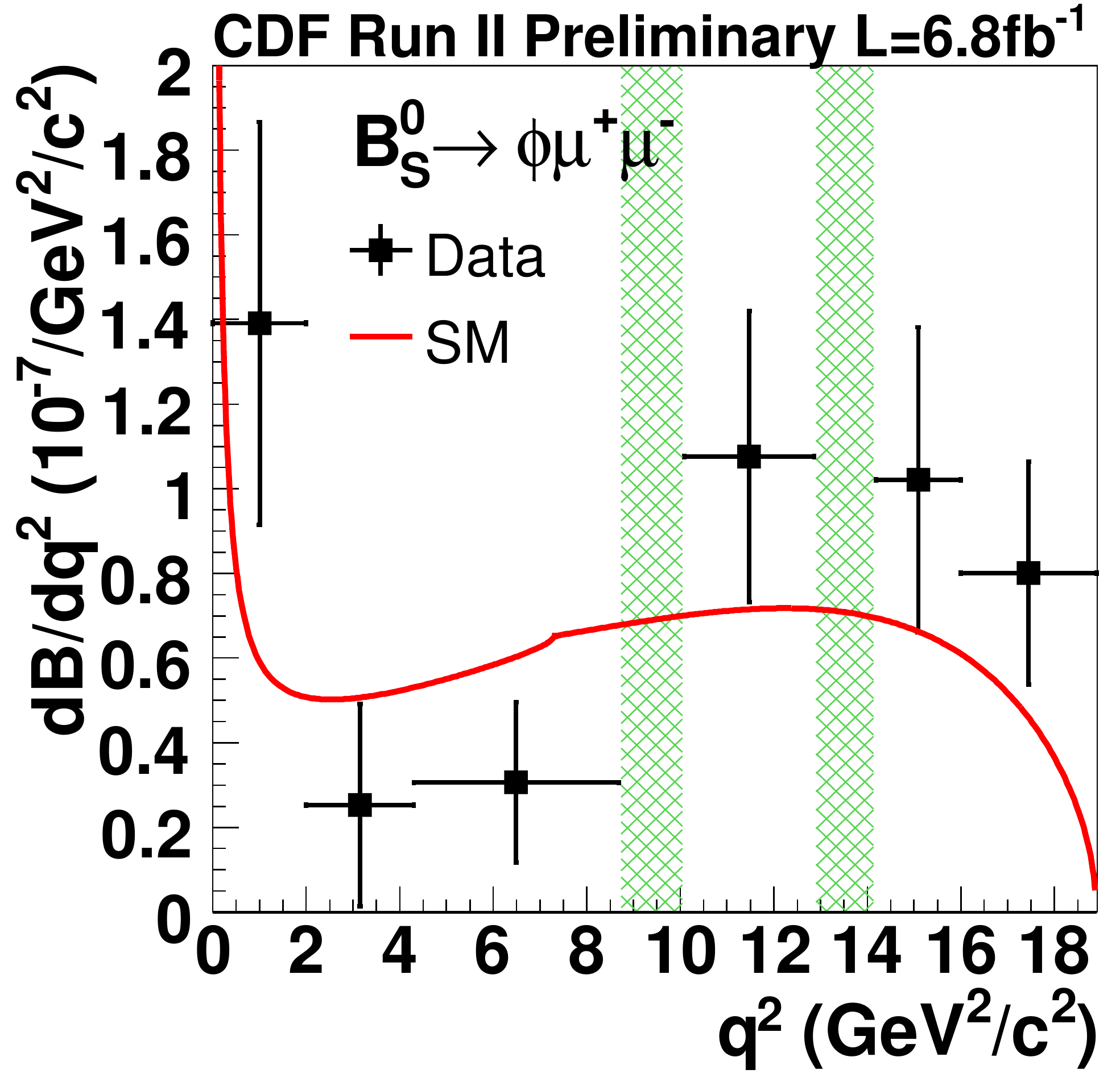}
\caption{The branching fraction for \( B_s^0 \to \phi \mu^+\mu^- \) as a function of \( q^2 \) of the dimuon system.
The green bands show the \( J/\psi \) and \( \psi(2S) \) veto regions.
A deviation of the measured branching fraction from the SM prediction would indicate the contribution to new physics to the decay.
}
\label{fig:Bstophimumu}
\end{figure}

Another recently observed flavor--changing neutral--current decay is \( B_s^0 \to \phi \mu^+\mu^- \).
An updated measurement using \( 6.8\,\mathrm{fb}^{-1} \) of integrated luminosity was released by CDF in Summer 2011 \cite{CDF_LbtoLmumu}.
The measurement of the branching fraction as a function of the \( q^2 \) of the dimuon system is displayed in Fig.~\ref{fig:Bstophimumu}, and is sensitive to new physics.

\subsubsection{ \boldmath \( B_s^0 \to D_s^{(*)-} D_s^{(*)+} \) \unboldmath}

It has been suggested that the majority of the width difference between the heavy and light \( B_s^0 \) states, \( B_{sH} \) and \( B_{sL} \), comes from the two body modes to a pair of \( D_s \) mesons, ground state or excited state.
The \( D_s^+ D_s^- \) state is \( CP \) even, and the final states with \( D_s D_s^* \) or \( D_s^* D_s^* \) mesons are assumed to be predominantly \( CP \) even.
The sum of the three branching fractions is, under the above assumption, approximately equal to \( \Delta \Gamma_s / \Gamma_s \).
More recent work has questioned the assumption that these modes dominate the width difference, but this is an issue that can be settled with additional measurements.

\begin{table}[t]
\begin{center}
\caption{The branching fraction measurements for \( B_s^0 \to D_s^{(*)-} D_s^{(*)+} \) and significance \(S\) from the Belle collaboration.}
\begin{tabular}{l @{\hspace{5mm}} c @{\hspace{5mm}} c}
\hline\hline  mode &  \( \mathcal{B} \) (\%) & \( S \)
\\ \hline
\( D_s^+ D_s^- \) & \( 0.58^{+0.11}_{-0.09} \pm 0.13 \) & \( 11.6 \) \\
\( D_s^{\ast +} D_s^- + D_s^{\ast -} D_s^+ \) & \( 1.8 \pm 0.2 \pm 0.4 \) & \( 13.3 \) \\
\( D_s^{*+} D_s^{*-} \) & \( 2.0 \pm 0.3 \pm 0.5 \) & \( 8.6 \) \\
Sum & \( 4.3 \pm 0.4 \pm 1.0 \) &  \\ \hline
\( \Delta \Gamma_s / \Gamma_s \) & \multicolumn{2}{c}{ \( (9.0 \pm 0.9 \pm 2.2 ) \% \) } \\
\hline\hline
\end{tabular}
\label{tab:BstoDsDs}
\end{center}
\end{table}

The Belle collaboration has succeeded in measuring all three branching fractions \cite{BELLE_BstoDsDs} obtaining the results listed in Table~\ref{tab:BstoDsDs}.
If these branching fractions dominate the difference between the widths of the heavy and light eignestates, then the sum of the branching fractions is related to the relative width difference by
\begin{equation}
\frac{\Delta\Gamma_s}{\Gamma_s} = \frac{2 \mathcal{B}(B_s^0 \to D_s^{(*)-} D_s^{(*)+}) } { 1 - \mathcal{B}(B_s^0 \to D_s^{(*)-} D_s^{(*)+}) }
\end{equation}
where \( \mathcal{B}(B_s^0 \to D_s^{(*)-} D_s^{(*)+}) \) is the sum of all three branching fractions.
The result of \( 9.0\% \) is consistent, within errors, with other averages for the relative decay width and theoretical expectations.
This measurement is promising to perform with more data, as well as to measure some of the other modes that are thought could contribute to the width difference.

\subsubsection{ \boldmath \( B_s^0 \to \pi^+\pi^- \) and \( B^0 \to K^+K^- \) \unboldmath}

The charmless, two--body decays of neutral \( B \) hadrons to charged pions and kaons are rich in physics.
The measurements of the branching fractions and decay asymmetries have produced numerous results.
All of the possible decay modes to charged pions and kaons were seen except two, \( B_s^0 \to \pi^+\pi^- \) and \( B^0 \to K^+K^- \).
These modes have the feature that none of the quarks in the initial state meson exist in the final state particles and are known as annihilation or penguin annihilation decays.
Their branching fractions are suppressed by more than an order of magnitude from the branching fraction to the favored two--body pion/kaon mode.

Earlier this year, the CDF experiment produced \( 3.7\sigma \) evidence for the \( B_s^0 \to \pi^+\pi^- \) decay \cite{CDF_Btohh}, followed by a \( 5.2\sigma \) observation by LHCb \cite{LHCb_Btohh}.
The branching fractions determined from their fits are
\begin{eqnarray*}
& \mathcal{B}(B_s^0 \to \pi^+\pi^-) = & \\
& [0.57 \pm 0.15 (\mathrm{stat}) \pm 0.10 (\mathrm{syst})] \times 10^{-6} & \mathrm{\ CDF} \\
& [0.98 \pm 0.21 (\mathrm{stat}) \pm 0.11 (\mathrm{syst})] \times 10^{-6} & \mathrm{\ LHCb.} \\
\end{eqnarray*}
These results are consistent within the experimental uncertainties.

Both experiments continue to have difficulty separating the \( B^0 \to K^+ K^- \) signal from nearby two--body modes, achieving about \( 2\sigma \) signal significance.
They produce branching fractions for this mode of
\begin{eqnarray*}
& \mathcal{B}(B^0 \to K^+K^-) = & \\
& [0.23 \pm 0.10 (\mathrm{stat}) \pm 0.10 (\mathrm{syst})] \times 10^{-6} & \mathrm{\ CDF} \\
& [0.14 \pm 0.06 (\mathrm{stat}) \pm 0.07 (\mathrm{syst})] \times 10^{-6} & \mathrm{\ LHCb} \\
\end{eqnarray*}
also in agreement within the uncertainties.

\subsubsection{\boldmath \( B_s^0 \to \Lambda_c^- \pi^+ \Lambda \) \unboldmath}

Belle reported the first observation of a baryonic decay of the \( B_s^0 \) meson.
Using \( 121\,\mathrm{fb}^{-1} \) of \( \Upsilon(5S) \) data, they reconstruct \( 24 \pm 7 \) decays \( B_s^0 \to \Lambda_c^- \pi^+ \Lambda \), where the baryons are reconstructed in the decays \( \Lambda_c^- \to \bar{p} K^+ \pi^- \) and \( \Lambda \to p \pi^- \) \cite{BELLE_BstoLcpiL}.
The branching fraction is determined to be
\begin{eqnarray*} 
& \mathcal{B}(B_s^0 \to \Lambda_c^- \pi^+ \Lambda) = & \\
& [4.8 \pm 1.4 (\mathrm{stat}) \pm 0.9 (\mathrm{syst}) \pm 1.3 (\Lambda_c^-) ] \times 10^{-4}. &
\end{eqnarray*}

\subsection{ \boldmath Triple Product Asymmetries in \( B_s^0 \to \phi \phi \) \unboldmath}

For the first time, triple product asymmetries for the decay \( B_s^0 \to \phi \phi \) with \( \phi \to K^+ K^- \).
The triple product asymmetries don't require flavor tagging, and are zero in the SM, making them advantageous to use with modest statistics.
There are two independent asymmetries, and the ones the experiments have chosen are defined as
\begin{eqnarray}
u & = & \sin\phi \cos\phi \\
v & = & \sin\phi \sign(\cos\theta_1 \cos\theta_2) \nonumber
\end{eqnarray}
where the \( \phi \) is the angle between the decay planes of the \( \phi \) mesons, and \( \theta_i \) are the angles between the \( B_s^0 \) boost direction and the direction of the \( K^+ \) in the helicity frame.

The CDF experiment has published the first measurement \cite{CDF_Bstophiphi}, with the results
\begin{eqnarray*}
A_u & = & -0.007 \pm 0.064 (\mathrm{stat}) \pm 0.018 (\mathrm{syst}) \\
A_v & = & -0.120 \pm 0.064 (\mathrm{stat}) \pm 0.016 (\mathrm{syst})
\end{eqnarray*}
The LHCb experiment produced a preliminary result \cite{LHCb_Bstophiphi}, with the results
\begin{eqnarray*}
A_u & = & -0.064 \pm 0.057 (\mathrm{stat}) \pm 0.014 (\mathrm{syst}) \\
A_v & = & -0.070 \pm 0.057 (\mathrm{stat}) \pm 0.014 (\mathrm{syst})
\end{eqnarray*}
The results are consistent with the SM expectation of zero and with each other, within the experimental uncertainties.
With more data, this decay can be analyzed in the same fashion as the \( J/\psi \phi \) decay, that is with a full angular analysis and flavor tagging, to contribute to the measurements made in the \( J/\psi \phi \) channel.

\subsection{\boldmath Search for the Rare Decays \( B^0_{(s)} \to \mu^+\mu^- \) \unboldmath}

The flavor--changing neutral--current decays \( B_s^0 \to \mu^+\mu^- \) and \( B^0 \to \mu^+\mu^- \) have been of particular interest due to their sensitivity to new physics.
In particular, the \( B_s^0 \) decay is extremely sensitive to supersymmetry, with a contribution proportional to \( \tan^6\beta \).
And it has been pointed out that, in the case that new physics is seen where the nature is not entirely known, the combination of these rare decays with other measurements ( \( b \to s\gamma \), \( B^+ \to \tau^+\nu \) and so forth) can help discriminate between various scenarios.

In the context of the SM, the branching fractions of these decays are estimated to be
\[ \mathcal{B}(B_s^0 \to \mu^+\mu^-) = ( 3.2 \pm 0.2 ) \times 10^{-9} \]
and
\[ \mathcal{B}(B^0 \to \mu^+\mu^-) = ( 1.0 \pm 0.1 ) \times 10^{-10}. \]
While experiments are still an order of magnitude away from the sensitivity needed to see the \( B^0 \) decay, they are now within a factor of 5 or 6 of the \( B_s^0 \) decay.
This is close enough to observe the decay if the branching fraction is increased by a factor of a few above the SM value by the presence of beyond the standard model physics.
And if there is no new physics increase, it seems likely that \( 3\sigma \) evidence for the long sought \( B_s^0 \) decay will occur in the near future.

\begin{table}[t]
\begin{center}
\caption{The 95\% CL limits for \(\mathcal{B}(B_{(s)}^0 \to \mu^+\mu^-) \).}
\begin{tabular}{l @{\hspace{5mm}} c @{\hspace{5mm}} c}
\hline\hline  &  \(B^0 \to \mu^+\mu^- \) & \(B_s^0 \to \mu^+\mu^- \)
\\ \hline
CDF & \( < 6.0 \times 10^{-9} \) & \( < 4.0 \times 10^{-8} \) \\
CMS & \( < 4.6 \times 10^{-9} \) & \( < 1.9 \times 10^{-8} \) \\
LHCb & \( < 5.2 \times 10^{-9} \) & \( < 1.5 \times 10^{-8} \) \\
Theory & \( (0.10 \pm 0.01) \times 10^{-9} \) & \( (0.32 \pm 0.02 ) \times 10^{-8} \) \\
\hline\hline
\end{tabular}
\label{tab:B0lims}
\end{center}
\end{table}

Three experiments reported new results for the search this summer:  CDF, CMS, and LHCb.
They also reported new limits for the \( B^0 \to \mu^+\mu^- \) decay, summarized in Table~\ref{tab:B0lims}.

\bigskip 
\begin{acknowledgments}
The authors wish to thank all the experiments for providing data on their measurements.

Work supported by the U.S. Department of Energy.
\end{acknowledgments}

\bigskip 
\bibliography{pic_harr}

\end{document}